\documentclass[prc,aps,superscriptaddress,twocolumn,showpacs,nofootinbib]{revtex4}

\usepackage{graphicx,amsmath,amssymb,bm,multirow}
\usepackage{amsthm}
\usepackage{amsfonts}

\usepackage{dcolumn}
\usepackage{bm}

\usepackage[usenames,dvipsnames]{color}

\newcommand{\be}[1]{\begin{equation}\label{#1}}
\newcommand{\ee}{\end{equation}}

\newcommand{\bra}[1]{\langle#1|}
\newcommand{\ket}[1]{|#1\rangle}
\newcommand{\braket}[2]{\langle#1|#2\rangle}
\newcommand{\matrEL}[3]{\langle#1|#2|#3\rangle}

\newcommand{\elemA}[2]{\ensuremath{{}^{#1}}\textrm{#2}}

\begin{document}
\title{Beyond-mean-field study of the possible ``bubble'' structure
       of \elemA{34}{Si}}

\author{Jiang-Ming~Yao}
\email{jmyao@swu.edu.cn}
\affiliation{Physique Nucl\'eaire Th\'eorique,
             Universit\'e Libre de Bruxelles, C.P. 229, B-1050 Bruxelles,
             Belgium}
\affiliation{School of Physical Science and Technology,
             Southwest University, Chongqing 400715, China}

\author{Simone~Baroni}
\email{simone.baroni@ulb.ac.be}
\affiliation{Physique Nucl\'eaire Th\'eorique,
             Universit\'e Libre de Bruxelles, C.P. 229, B-1050 Bruxelles,
             Belgium}

\author{Michael~Bender}
\email{bender@cenbg.in2p3.fr}
\affiliation{Universit\'e Bordeaux, Centre d'Etudes Nucl\'eaires de
            Bordeaux Gradignan, UMR5797, F-33175 Gradignan, France}
\affiliation{CNRS/IN2P3, Centre d'Etudes Nucl\'eaires de Bordeaux
            Gradignan, UMR5797, F-33175 Gradignan, France}

\author{Paul-Henri~Heenen}
\email{phheenen@ulb.ac.be}
\affiliation{Physique Nucl\'eaire Th\'eorique,
             Universit\'e Libre de Bruxelles, C.P. 229, B-1050 Bruxelles,
             Belgium}

%
%

\begin{abstract}

Recent self-consistent mean-field calculations predict a substantial
depletion of the proton density in the interior of $\elemA{34}{Si}$.
In the present study, we investigate how correlations beyond the mean field modify this finding. The framework of the calculation is a particle-number and angular-momentum projected Generator Coordinate Method based on
Hartree-Fock-Bogoliubov+Lipkin-Nogami states with axial quadrupole deformation.
 The parametrization
SLy4 of the Skyrme energy density functional is used together with a density-dependent pairing energy functional. For the first time, the generator coordinate method is applied to the calculation of charge and transition densities. The impact of pairing
correlations, symmetry restorations and shape mixing on the density profile is analyzed step by step. All these effects significantly alter the radial
density profile, and tend to bring it closer to a Fermi-type density distribution.

\end{abstract}



\maketitle
%
%
\section{Introduction}
\label{intro}

Charge distributions in atomic nuclei
\cite{Uberall71,Barrett74,Friar75,Sick85,Friedrich86,Frois87,Hodgson92}
provide very detailed information about nuclear structure.
They are obtained through the analysis of elastic electron-nucleus scattering
data. Because of the absence of suitable targets, data on unstable nuclei are available, to the best of our knowledge, only for $\elemA{14}{C}$~\cite{Kline73} and $\elemA{3}{H}$~\cite{Beck87,Amrouna94}.
The SCRIT project~\cite{Wakasugi04,Suda05,Suda09} of constructing a
high-resolution electron spectrometer that is underway in Japan and
ELISe~\cite{Antonov11}, planned to be constructed at FAIR, are expected
to provide data about the charge distributions and transition form factors
for many exotic nuclei in the future.

Because of the saturation properties of the nuclear medium, the radial dependence of the nuclear density takes, at the lowest order, the form of a Fermi distribution. However, the density often deviates from this simple behavior because of quantal effects related to the filling of single-particle states with wave functions that have specific spatial behavior. In this context, $s_{1/2}$ orbits in spherical nuclei have a very peculiar signature, as they are the only ones that contribute to the density at the nuclear center. Depending on whether they are filled or empty, $s_{1/2}$ orbits can generate a central bump in the density as it has been observed for $\elemA{40}{Ca}$~\cite{Sick79}, or a
central depression.

Mean-field-based methods~\cite{Bender03} are the tools of choice when modeling the nuclear density distribution. Indeed, they include the ingredients required for this
task: the full model space of occupied single-particle states as degrees of freedom together
with an effective interaction that reproduces the empirical saturation
properties of nuclear matter.

The density profile and the spatial dependence of the
single-particle potentials are closely related and self-consistently linked to each other. A central depression in the density might be accompanied by two very specific properties of the mean-field potential, one related to its central part and a second one related
to the spin-orbit potential.

A central depression of the density is reflected in the central potential by a maximum at the origin and a minimum for some finite distance $r$. This is often called a ``wine-bottle'' shaped central potential, referring to the shape of the bottom of a bottle of wine. Levels with low orbital angular momentum $\ell$ are then
pushed up relatively to those with large $\ell$ that are pulled down.
For sufficiently large rearrangement, the order of single-particle
levels can even change, lowering the central density even more
and leading to the so-called ``bubble nuclei''. For specific
``bubble magic numbers'' 18, 34, 50, 58, 80, 120, \ldots \cite{Wong72,Wong73}
large shell effects might compensate for the loss in binding
energy due to the reduced central density well below the nuclear matter saturation value. There was some speculation in the 1970s whether such structure could exist in nuclei that were about to become accessible for detailed studies, in particular $\elemA{36}{Ar}$ and some Hg isotopes
\cite{Wong72,Wong73,Davies72,Davies73,Campi73,Beiner73,Nilsson74,Saunier74}.
However, the possibility of a bubble structure in these nuclei has been ruled out by experiment. By contrast, predictions that superheavy and hyperheavy nuclei beyond the currently known  region of the mass table might take the form of bubbles
\cite{Dietrich97,Decharge97,Bender99,Decharge03,Nazarewicz00} are still standing. In fact, for a large charge number $Z$, a hollow density distribution is energetically favored over a regular one as it lowers the Coulomb repulsion. In this context, one often distinguishes between ``true bubbles'', which have vanishing density in their center, and
``semi-bubbles'', which have a central density significantly lower than saturation density, but with a non-zero value.

The second effect of a central depression concerns the spin-orbit potential.
In self-consistent mean-field models, this potential is proportional to the gradient of a combination of proton and neutron densities, whose relative weights depend on the model and parametrization
\cite{Bender03,Bender99}. For nuclei with a regular density profile, it is peaked at the nuclear surface. For nuclei with a central depletion of the density, the spin-orbit potential has a second peak of opposite sign in the nuclear interior. This usually reduces the spin-orbit splitting of orbits located mainly at the nucleus center, whereas that of orbits situated at the
nuclear surface is not affected.

Recently, there has been a renewal of interest in nuclei presenting
a hollow in their density distributions. Some modern parametrizations
of the relativistic mean field \cite{Todd-Rutel2004,Chu2010} and of
the Skyrme energy density functional (EDF) \cite{Khan2008,Grasso2009}
predict a hollow proton density for $\elemA{34}{Si}$ and some neutron-rich
Ar isotopes. At the time being, $\elemA{34}{Si}$ stands out
as the only candidate on which many different effective interactions
agree. The possible proton bubble structure of this nucleus has also been suggested as an explanation of the results on the transfer reactions $\elemA{36}{S}(d,p)\elemA{37}{S}$ and $\elemA{34}{Si}(d,p)\elemA{35}{Si}$. Indeed, the splitting between the observed  \emph{neutron} $3/2^-$ and $1/2^-$ levels that have the largest spectroscopic factors in the $2p$ shell is decreased from $\elemA{37}{S}$  ($\approx 1.7$ MeV) to $\elemA{35}{Si}$
  ($\approx 1.1$ MeV)
 \cite{Burgunder2011,Sorlin2011}.


Besides the debatable interaction dependence of the density
distributions, one may wonder whether bubble-type structures
are stable against correlation effects. Indeed, any correlation will
inevitably populate empty levels and in particular the $2s_{1/2}$, even
in models like the one that we use here, where its population
cannot be easily singled out. In fact, it is known for a long time that,
already for nuclei with a more regular density distribution, mean-field
calculations tend to overestimate the spatial fluctuations of the density
when compared to data. Correlations usually tend to flatten out the density
distribution, and often bring it closer to data. The effect of
pairing has been studied in Ref.~\cite{Bennour89}, and the impact of
fluctuations in shape degrees of freedom has been studied within the
random phase approximation (RPA) for many spherical nuclei
\cite{Faessler76,Reinhard79,Gogny79,Decharge83,Esbensen83,Johnson88,
Barranco87,Sil08} and using a one-dimensional~\cite{Girod76} or
five-dimensional~\cite{Gogny79,Decharge83,Girod82} microscopic collective
Hamiltonian for some transitional ones.

The most obvious correlations that could reduce the central
depression of the density are due to pairing \cite{Beiner73}. However, many calculations made for bubble nuclei do not include them~\cite{Davies72,Davies73,Campi73,Saunier74}.
This can be justified for $\elemA{34}{Si}$,
where the large gap between the proton $1d_{5/2}$ and $2s_{1/2}$ levels suppresses pairing, resulting in its unphysical collapse when treated with the Hartree-Fock-Bogoliubov (HFB) method.
Pairing correlations have then to be treated beyond the mean field approximation.
Another kind of correlations that might affect the density
profile is related to the spreading of the ground-state wave function
around the mean-field configuration. It has been pointed out in
Ref.~\cite{Bender06a} that the ground states of most light nuclei
 may show strong shape fluctuations that in general lead
to a substantial increase of their charge radii when treated in a
beyond-mean-field framework. The same effect may also strongly
influence the density profile.

In the following, we will compare results obtained from
calculations that successively add correlations to the ground-state wave function:
\begin{enumerate}
\item[(i)]
spherical mean-field calculations without taking
pairing correlations into account (HF)
\item[(ii)]
spherical mean-field calculations including
pairing correlations within the HFB+Lipkin-Nogami (HFB+LN)
scheme, which constitutes an approximate variation after projection
on particle number
\item[(iii)]
particle-number projection after variation of the spherical
mean-field state obtained in HFB+LN
\item[(iv)]
configuration mixing of angular-momentum $J = 0$ and particle-number
projected mean-field states with different intrinsic axial quadrupole
moment. We will refer to these wave functions in the following as
symmetry-restored generator coordinate method (GCM).
\end{enumerate}
In addition, we will study how much the observable charge density,
obtained through convolution of the proton density with the proton's
internal charge distribution, differs from the point proton density
used to calculate the energy.

Symmetry-restored GCM has been used to describe ground-state
correlations and collective excitation spectra of a large range
of nuclei with reasonable success
\cite{Egido04,Bender08LH,Niksic06,Bender2008,Rodriguez10,Yao10}.
Its actual implementations are not yet flexible
enough to reproduce correctly all details of excitation spectra,
mainly because of the usually too low moment of inertia.
However, this method describes rather well properties related to the
nuclear shape, such as transition probabilities. It is therefore
 important to enlarge its range of applications by the calculation of charge and transition
densities in the laboratory frame. In this paper, we report on
a first application that enables us to illustrate the effect of various
kinds of correlations on the density distribution of $\elemA{34}{Si}$.

 We will first give a brief outline of the model.
Results for low-lying collective states in
$\elemA{34}{Si}$ are presented in Sect.~\ref{results_spectroscopy},
whereas the modification of the ground state density distribution
brought by correlations is  discussed in Sect.~\ref{results_density}.
Section~\ref{concl} will summarize our findings.


\section{Calculational details}
\label{calc}

The self-consistent HFB equations are solved on a cubic three-dimensional
coordinate-space mesh extending from $-11.2$~fm to 11.2~fm in each direction
with a step size of 0.8~fm. Thanks to the reflection symmetry with
respect to the $x=0$, $y=0$ and $z=0$ planes imposed on the single-particle
wave functions in our code~\cite{Bonche05}, it is sufficient to
solve the HFB equations in 1/8 of the box. The HFB equations are
complemented by the Lipkin-Nogami prescription to avoid the unphysical
breakdown of pairing correlations at low level density. A constraint on
the axial mass quadrupole moment
$q \equiv \langle Q_2 \rangle = \langle 2 z^2 - x^2 - y^2 \rangle$
is used to construct mean-field states $| q \rangle$ with different
intrinsic deformation.

Eigenstates of the particle-number operators $\hat{N}$ and $\hat{Z}$
are obtained by applying the projection operator
\begin{equation}
\label{proj_N}
\hat{P}_{N_0}
= \frac{1}{2 \pi}
  \int_{0}^{2 \pi} \! \! d\varphi \; e^{i \varphi (\hat{N}-N_0)}
\end{equation}
for neutrons and protons. Eigenstates of the total
angular momentum in the laboratory frame $\hat{J}^2$ and its $z$
component $\hat{J}_z$ with eigenvalues $\hbar^2 J (J+1)$ and $\hbar M$,
respectively, are obtained by applying the operator
\begin{equation}
\label{proj_J}
\hat{P}^J_{MK}
= \frac{2J+1}{8\pi^2} \!
  \int_0^{2\pi} \! \! d\alpha \int_0^\pi \! \! d\beta \, \sin(\beta)
  \int_0^{2\pi} \! \! d\gamma \, \mathcal{D}^{J*}_{MK} \, \hat{R}
\end{equation}
that contains the rotation operator $\hat{R}
= e^{-i\alpha \hat{J}_x} \, e^{-i\beta \hat{J}_y} \, e^{-i\gamma \hat{J}_z}$
and the Wigner rotation matrix $\mathcal{D}^{J}_{MK} (\alpha,\beta,\gamma)$
on the nucleus' wave function. Both depend on the Euler angles $\alpha$,
$\beta$ and $\gamma$. The operator $\hat{P}^J_{MK}$ picks the component
with projection $K$ along the intrinsic $z$ axis from the mean-field state.
Throughout this study, we will restrict ourselves to axial states with
$K=0$. As angular-momentum projected states will be always
projected also on particle-number, we drop the indices $N_0$ and $Z_0$for the sake of notation:
\begin{equation}
\label{eq_GCM:08}
\ket{J M q}
= \frac{\hat{P}^J_{M0} \hat{P}_{N_0} \hat{P}_{Z_0} \ket{q}}
       {\sqrt{\bra{q} \hat{P}^J_{00} \hat{P}_{N_0} \hat{P}_{Z_0} \ket{q} }}
\, .
\end{equation}
GCM~\cite{Ring80} is a very flexible tool that in particular allows
us to study the spreading of the mean-field ground-state wave
function in collective degrees of freedom. It will be
used here to study the fluctuations of the spherical ground state of
$\elemA{34}{Si}$ with respect to the axial quadrupole moment assuming
a superposition of projected HFB+LN states of different deformation
$| q \rangle$:
\begin{equation}
\label{eq_GCM:10}
\ket{J M \mu}
= \sum_q f_\mu^{J} (q) \, \ket{J M q}
\, .
\end{equation}
The weight factors $f_\mu^{J}(q)$ and the energies $E_\mu^{J}$
of the states $\ket{J M \mu}$ are the solutions of the
Hill-Wheeler-Griffin equations~\cite{Ring80}
\begin{equation}
\label{eq_GCM:20}
\sum_{q'} \left[ \mathcal{H}^{J}(q,q') - E_\mu^{J}\mathcal{N}^{J}(q,q')
          \right] \, f_\mu^{J}(q')
= 0
\, ,
\end{equation}
where the norm kernel reads
$\mathcal{N}^{J}(q,q') = \braket{J M q}{J M q' }$ and where the energy
kernel is given by a multi-reference energy density functional that
depends on the mixed density matrix between the two projected states
$\ket{J M q}$ and $\ket{J M q' }$ \cite{Lacroix09}.

Throughout this article, we will use the parametrization SLy4
\cite{Chabanat98} of the Skyrme energy density functional together
with a local pairing energy functional of surface type \cite{Rigollet99}
with parameters $\rho_0 = 0.16$ fm$^{-3}$ for the switching density and
$V_{0} = -1000.0$ MeV fm$^3$ for the pairing strength unless noted
otherwise. A soft cutoff at $\pm 5$ MeV around the Fermi energy is used
when solving the HFB equations as described in Ref.~\cite{Rigollet99}.
More details about the calculations of the GCM kernels can be
found in Ref.~\cite{Bender2008} and references given therein.

The weight functions $f^{J}_\mu(q)$ in Eq.~(\ref{eq_GCM:10}) are not
orthogonal. A set of orthonormal collective wave functions $g_\mu^{J}(q)$
can be constructed as~\cite{Ring80}
\begin{equation}
\label{eq_GCM:30}
g_\mu^{J}(q)
= \sum_{q'} \big( \mathcal{N}^{J}\big)^{1/2}(q,q')  \, f_\mu^{J}(q')
\, .
\end{equation}
It has to be stressed, however, that $|g_\mu^{J}(q)|^2$ does not represent
the probability to find the deformation $q$ in the GCM state $\ket{J M \mu}$.
In addition, in the absence of a metric in the definition of the correlated
state $\ket{J M \mu}$, Eq.~(\ref{eq_GCM:10}), the values of $g_\mu^{J}(q)$
for a converged GCM solution still depend on the discretization chosen for
the collective variable $q$, which is not the case for observables like the
energy or transition probabilities.

The spatial density distribution of the projected GCM states is
constructed as the expectation value of the operator
$\hat{\rho}(\vec{r}) = \sum_i^A ( \hat{\vec{r}}-\vec{r}_i)$,
\begin{eqnarray}
\label{eq_GCM:70}
  \lefteqn{    \rho^{JM \mu}(\vec{r}) =
              \matrEL{ JM \mu }{ \; \hat{\rho}(\vec{r}) \; }{ JM \mu }
          }
  \nonumber \\
  & = & \sum_{qq'} f_\mu^{J*}(q) \,
                  \matrEL{JM q } { \; \hat{\rho}(\vec{r}) \; } {JM q'}
         \, f_\mu^{J}(q')
  \nonumber \\
  & = & \sum_{qq'}
        \dfrac{f_\mu^{J*}(q) f_\mu^{J}(q')}
              {\sqrt{\bra{q} \hat{P}^J_{00} \hat{P}_{N_0} \hat{P}_{Z_0} \ket{q}}
               \sqrt{\bra{q'} \hat{P}^J_{00} \hat{P}_{N_0} \hat{P}_{Z_0} \ket{q'}}} \nonumber \\
        &   & \times
        \dfrac{2J+1}{8\pi^2}
        \int d\Omega^\prime \mathcal{D}^{J\ast}_{0 M}(\Omega^\prime) \,
              \sum_{K}\mathcal{D}^{J}_{KM}(\Omega^\prime) \nonumber \\
        &   & \times\hat
              R^{\dag}(\Omega^\prime) \dfrac{(2J+1)}{2}
              \int^\pi_0 \! \! d\beta\sin(\beta) \,
              d^{J}_{K0} (\beta) \nonumber \\
        &   & \times \langle q \vert \hat R(\beta)
              \sum_{i}\delta(\vec{r}-\vec{r}_i) \,
              \hat P_{N_0}\hat P_{Z_0}  \vert q' \rangle \, ,
\end{eqnarray}
where we use the shorthand $\Omega \equiv (\alpha, \beta,\gamma)$
for the Euler angles. Note that the calculation of the density
in the laboratory frame requires projectors on the left and on the right.
More details on the calculation of the correlated density will be given
in a forthcoming publication~\cite{Yao2012}.


\section{Spectroscopy of low-lying states}
\label{results_spectroscopy}

\begin{figure}[t!]
\begin{center}
\includegraphics[clip=,width=0.50\textwidth]{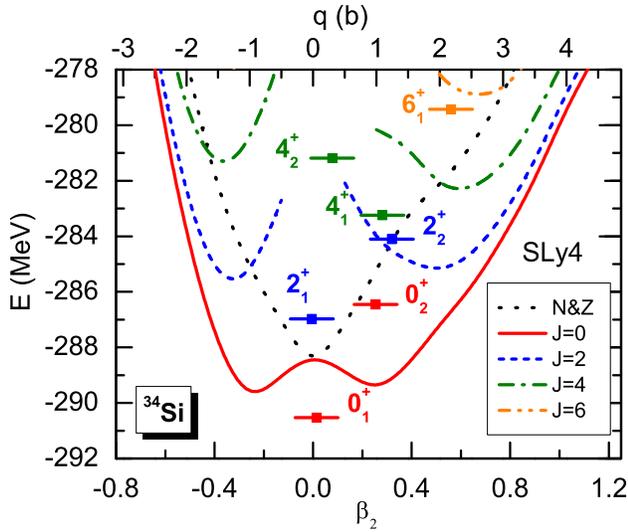}
\end{center}
\vskip-1.2cm
\caption{\label{energy_curves}
(Color online)
Energy curves for the particle-number-projected HFB states
(N\&Z) and particle-number and angular-momentum projected states
($J=0$, 2, 4, 6 curves) for $\elemA{34}{Si}$ as a function of
the intrinsic quadrupole deformation of the mean-field states they are
projected from. The solid square dots correspond to the lowest GCM
solutions, which are plotted at their average deformation
$\sum_q q \; |g_\mu^{J}(q)|^2$ (see text).
}
\end{figure}

The energy curves obtained after projection on particle number
and on angular momentum are shown in Fig.~\ref{energy_curves}.
The abscissa corresponds to the mass quadrupole moment $q$ of the
intrinsic states that is projected (upper scale) and, equivalently,
to the dimensionless quadrupole deformation
\begin{equation}\label{beta_q}
\beta_2
= \sqrt{\frac{5}{16\pi}}\frac{4\pi}{3R^2A} \, \langle Q_2 \rangle
\, ,
\end{equation}
where $R = 1.2 \, A^{1/3}$ fm.

\begin{figure}[t!]
\begin{center}
\includegraphics[clip=,width=0.50\textwidth]{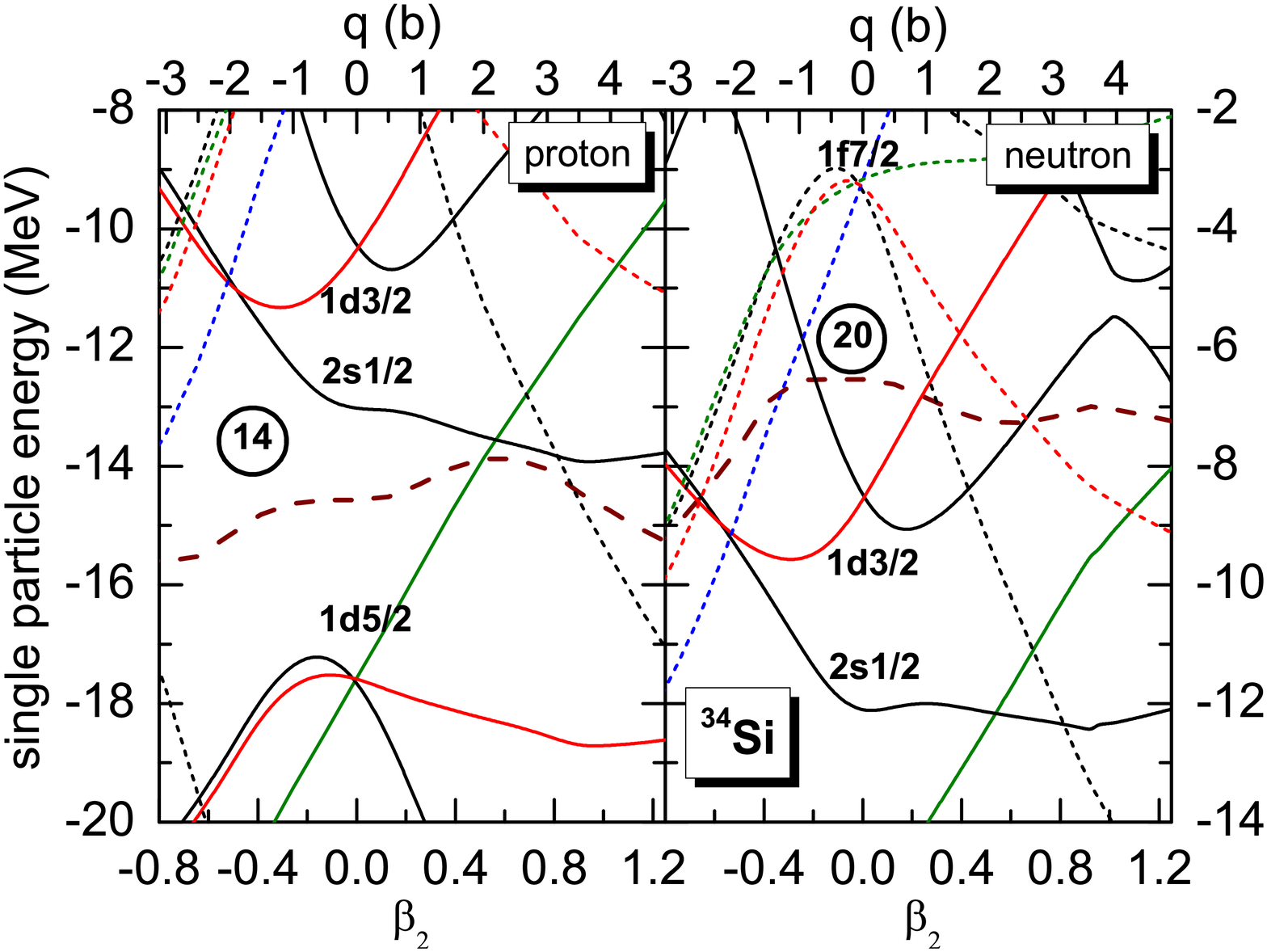}
\end{center}
\vskip-0.6cm
\caption{ \label{spe}
(Color online)
Nilsson diagram of the eigenvalues of the single-particle Hamiltonian
for neutrons (left panel) and protons (right panel) as obtained with
the Skyrme interaction SLy4 for $\elemA{34}{Si}$ as a function of the
quadrupole deformation. Solid (dotted) lines represent levels of positive
(negative) parity, and black, red, green and blue color represents
levels with expectation values of $\langle j_z \rangle = 1/2$, $3/2$, $5/2$
and $7/2$. The thick long-dashed line represents the Fermi energy.
Single-particle levels are labeled for the spherical configuration only.
}
\end{figure}

The particle-number-projected energy curve presents a spherical
minimum with a steep rise with deformation (dotted line), as expected
for a nucleus with large neutron and proton shell gaps, cf.\ Fig.~\ref{spe}.
The projection on total angular momentum $J=0$ leads to
energy curves with prolate and oblate minima at about the same
deformation $|\beta_2| \approx 0.26$. The presence of these two minima is a usual result of angular-momentum projection when the  non-projected energy curve presents a spherical minimum~\cite{Bender06a,Egido04,Bender08LH}.
In fact, at small deformation, the states of a given $J$ projected out from prolate and oblate mean-field states with the same $| \beta_2 |$ value are almost equivalent. Oblate and prolate minima are also obtained for higher $J$-values. Our results are very similar to those of a similar calculation using the Gogny interaction~\cite{Rod2000a}.

\begin{figure}[b!]
\begin{center}
\includegraphics[clip=,width=0.50\textwidth]{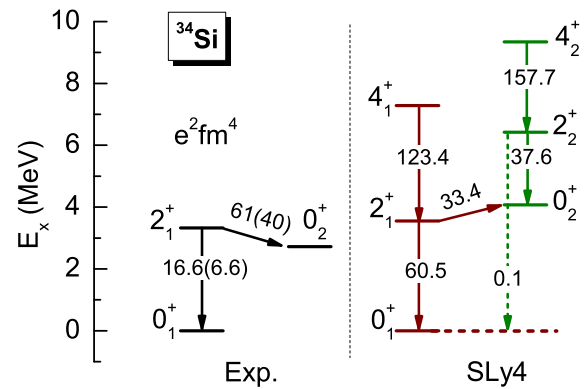}
\end{center}
\vskip-1.2cm
\caption{(Color online) Comparison between the experimental (left)
and calculated (right)  excitation energies $E_x$ and $B(E2)$ values
(in units of $e^2$ fm$^4$) for the low-lying states of $\elemA{34}{Si}$.
Experimental data are taken from Ref.~\cite{Rotaru2011}.}
\label{spectrum}
\end{figure}

The energies $E_\mu^{J}$ of the lowest GCM states are also shown
in Fig.~\ref{energy_curves} (solid square with a label $J^\pi_\mu$)
at the mean deformation $\sum_q q \; |g_\mu^{J}(q)|^2$ of the mean-field
states on which they are build. This mean deformation is
not an observable; still, it often provides a good indication about
the dominating mean-field configurations in a GCM state. The mean
deformations and the $B(E2)$ transition strengths suggest
to organize the correlated states into the two structures displayed
in Fig.~\ref{spectrum}, where the computed transition probabilities and
the energy of the levels are also compared with the available
experimental values~\cite{Rotaru2011}. Our result can be interpreted as
resulting from the coexistence of an anharmonic spherical vibrator and
a prolate deformed band at low excitation energy.  Both structures are
not pure and distorted by their strong mixing.

The energy of the recently observed low-energy $0^+_2$ state~\cite{Rotaru2011}
and the out-of-band $B(E2:2^+_1 \rightarrow 0^+_2)$ value are reproduced
rather well. However, the electric monopole $\rho^2(E0;0^+_2\rightarrow0^+_1)$
and the in-band $B(E2;2^+_1 \rightarrow 0^+_1)$ are overestimated by our
model: $58.1 \times 10^{-3}$ compared to the experimental value
of $13.0\, (0.9) \times 10^{-3}$~\cite{Rotaru2011} for the
former and 60.5 $e^2$ fm$^4$ compared to 16.6 $e^2$ fm$^4$ for
the latter. This discrepancy might indicate~\cite{Heyde88} that
the two lowest $0^+$ GCM states are too strongly mixed in our calculation.
The corresponding collective wave functions $g_\mu^{J}(q)$ are displayed
in Fig.~\ref{wave_functions}. Both are indeed spread over a very wide
range of deformations, with similar contributions at small deformation
$| \beta_2 | \approx 0$. The ground state is peaked around the deformations
of the two minima in the $J=0$ projected energy curve, cf.\
Fig.~\ref{energy_curves}. By contrast,the wave function of the
$0^+_2$ state is peaked at large prolate and oblate deformations where
at least one downsloping level from the neutron $f_{7/2}$ shell becomes
intruder by crossing the upsloping levels from the $sd$ shell, cf.\
Fig.~\ref{spe}. This is consistent with the interpretation of the
$0^+_2$ state in $\elemA{34}{Si}$ as a counterpart of the deformed
ground state of the slightly lighter nuclei located in the so-called
``island of inversion''~\cite{Rotaru2011}.


\section{Density distribution}
\label{results_density}

\begin{figure}[t!]
\begin{center}
\includegraphics[clip=,width=0.40\textwidth]{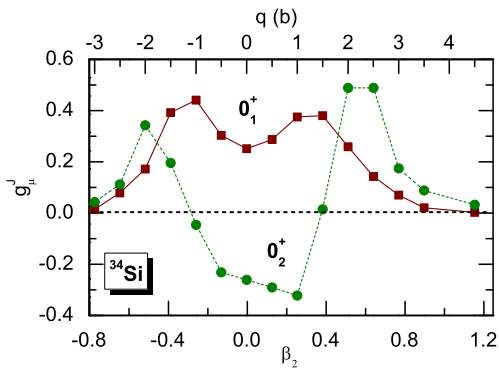}
\end{center}
\vskip-0.8cm
\caption{\label{wave_functions}
(Color online)
Collective wave functions $g_\mu^{J}(q)$ (cf.\ Eq.~(\ref{eq_GCM:30}))
of the two lowest GCM solutions for $J=0$.
}
\end{figure}

To quantify the depletion of the proton density distribution,
we will use a depletion factor
\begin{equation}
\label{eq:F}
F_{\text{max}}
\equiv \frac{\rho_{\textrm{max},p}-\rho_{\textrm{cent},p}}
            {\rho_{\textrm{max},p}}
\, ,
\end{equation}
which has been used in Ref.~\cite{Chu2010,Grasso2009} and
that measures the reduction of the density at the nucleus
center relatively to its maximum value.

\begin{figure}[b!]
\begin{center}
\includegraphics[clip=,width=0.47\textwidth]{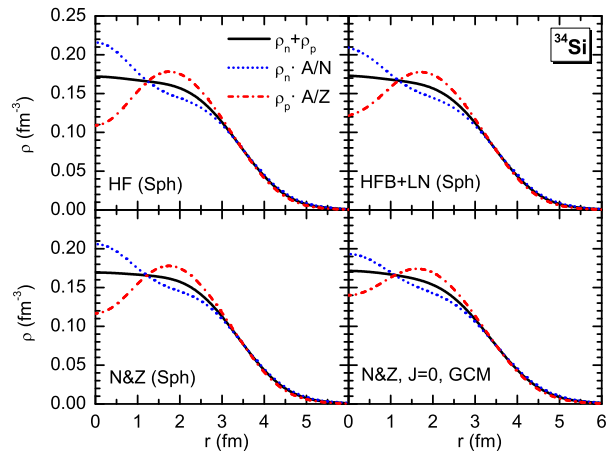}
\end{center}
\vskip-0.8cm
\caption{\label{scaled-dens}
(Color online) Neutron, proton and total radial densities
at $x = y = 0$ for $\elemA{34}{Si}$ for the spherical
HF state (top left panel), the spherical HFB+LN state (top right)
and its projection on good particle numbers ($N\&Z$, bottom left),
as well as for the GCM $0^{+}_1$ ground state ($N\&Z$, $J=0$, GCM, bottom
right). Neutron and proton densities have been rescaled with the
factors $A/N$ and $A/Z$, respectively.
}
\end{figure}

The effect of pairing correlations, projection on good quantum
numbers and configuration mixing on the radial profiles of the proton,
neutron and total densities is displayed in Fig.~\ref{scaled-dens}.
The densities of the HF, HFB+LN and particle-number projected HFB+LN
states are compared to those of the GCM $0^+$ ground state.
To facilitate the comparison, the proton and neutron densities
are rescaled by $A/Z$ and $A/N$ factors, respectively.

A large depletion at $r=0$ and a bulge at $r \approx 1.8$~fm are
obtained for the proton density when the HF method is used
(top left panel of Fig.~\ref{scaled-dens}). The HF neutron density,
however, has an opposite behavior, with a flat shoulder at intermediate
$r$ values and a bump at the nucleus center. This bump is similar to
the one found experimentally for the charge density in
$\elemA{40}{Ca}$~\cite{Sick79}.  Altogether, the total density has an
almost flat, even slowly rising, profile in the interior of the nucleus.
The same compensation of neutron and proton densities in the system's
interior is also found at all other stages of the calculation.

\begin{figure}[t!]
\begin{center}
\includegraphics[clip=,width=0.50\textwidth]{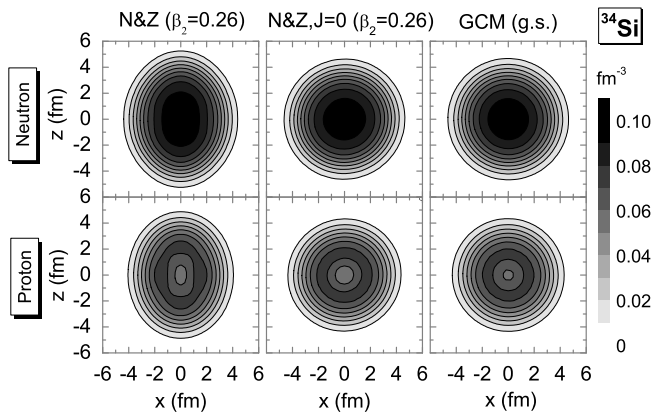}
\end{center}
\vskip-0.8cm
\caption{\label{contour_plot}
Contour plots of the neutron (upper panels) and proton (lower panels)
densities in the $y=0$ plane for the particle number projected HFB+LN
state with $\beta_2 = 0.26$ (left column), its projection on both particle
numbers and total angular momentum $J=0$ (middle column) and for the
$0^{+}$ GCM ground state (right column).
}
\end{figure}

Unconstrained HFB calculations for $\elemA{34}{Si}$ give the same
result as the HF approximation. This is because the large $Z = 14$ gap
of about 4.5~MeV between the proton $1d_{5/2}$ and $2s_{1/2}$ levels
in the single-particle spectrum prevents the protons from becoming
superfluid at the HFB approximation. The even larger $N = 20$ gap in
the single-particle spectrum has the same effect for neutrons.
The situation is different for nuclei such as $^{22}$O and $^{46}$Ar,
where pairing correlations are already active at the HFB level and wash
out the bubble structure predicted by HF calculations~\cite{Grasso2009}.

The collapse of pairing correlations when the density of single-particle
levels falls below a critical value is a deficiency of the HFB
method~\cite{Ring80,Rodriguez05}. It can be partially
corrected by using the LN procedure (top right panel of
Fig.~\ref{scaled-dens}). The level occupation is then smeared over
the Fermi energy and the proton $2s_{1/2}$ orbital becomes partially
occupied. As a consequence, the central proton density rises considerably
from  $F_{\text{max}} = 0.41$ (HF) to $F_{\text{max}} = 0.32$
(HFB+LN). The HFB+LN density presented in Fig.~\ref{scaled-dens}
is calculated using occupation
numbers corrected for particle-number projection by the approximation described for example in Ref.~\cite{Quentin90}. Using the
non-corrected BCS occupation numbers instead would overestimate the
effect of pairing and give a much larger reduction of the depletion
factor.

Projection of the HFB+LN state on good particle numbers (bottom left
panel of Fig.~\ref{scaled-dens}) substantially reduces the pairing
correlations, and the density profiles almost go back to the
HF ones with $F_{\text{max}}=0.36$. This reflects the well-known fact
that the LN approximation overestimates the correlations in the weak
pairing limit  (whereas HFB underestimates them), cf.\ for example
Ref.~\cite{Anguiano02}, and indicates that in this case a correct
treatment of pairing requires to go beyond the mean field.

The behavior of the density close to the origin is usually discussed
in terms of the occupation of single-particle states in the spherical
HF basis. This is not obvious in a method like the one that we use where
the mean-field basis is different for each deformation. Deformation
mixes single-particle states with different orbital angular
momentum. In particular, when one expands a deformed basis in terms
of the spherical one, the proton $2s_{1/2}$ level gets partially filled.
The situation is even more complicated after projection and
configuration mixing, cf.\ the bottom right panel of
Fig.~\ref{scaled-dens}. We have seen in Fig.~\ref{wave_functions}
that the collective wave function of the $0^+$ GCM ground state is
spread over a wide range of intrinsic deformations.

Figure~\ref{contour_plot} illustrates
how the density distribution of neutrons (upper panels) and protons
(lower panels) is modified at different levels of our calculation.
The left column shows contour plots of both densities for the
particle-number projected HFB+LN state with $\beta_2 = 0.26$ that
after angular-momentum projection gives the prolate minimum of the
$J=0$ curve in Fig.~\ref{energy_curves}. The proton density still
exhibits a central depletion, but  less pronounced than it is for the spherical
HFB+LN state, reflecting the partial filling of the $2s_{1/2}$ level
by deformation. After projection on total angular momentum $J = 0$
(middle column), the density is obtained in the laboratory frame and is spherical.
However, the central depression of the density is similar to the one
found when projecting on particle numbers only.
The configuration mixing leading to
the GCM $0^+$ ground state (right column) increases the central
proton density again and simultaneously reduces the value at the bulge, see the bottom right panel of Fig.~\ref{scaled-dens}, which in
combination reduces the depletion factor to $F_{\text{max}} = 0.21$.
The values of central and maximum densities and of the depletion
factor for the states discussed above are summarized in
Table~\ref{tab:F:proton}.

\begin{figure}[b!]
\begin{center}
\includegraphics[clip=,width=0.47\textwidth]{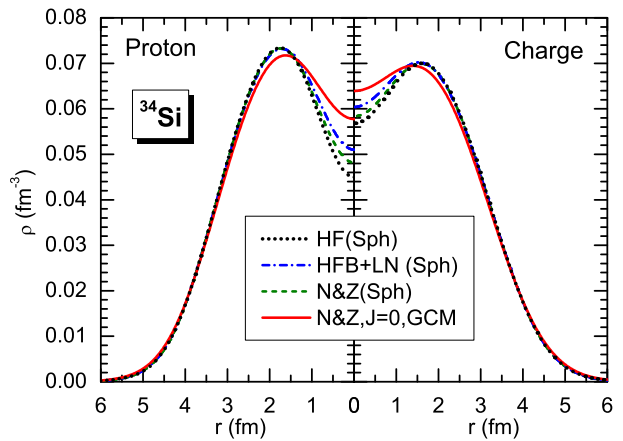}
\end{center}
\vskip-0.8cm
\caption{
(Color online)
Comparison of point-proton densities (left) with the folded charge
densities (right) for $\elemA{34}{Si}$ for the same states as
in Fig.~\ref{scaled-dens}.
}
\label{charge-dens}
\end{figure}

Up to now we have discussed the density of point protons and neutrons,
thereby neglecting that protons and neutrons are composite particles of extended size.
When searching for experimental signature of a depleted central density
in $\elemA{34}{Si}$ by elastic electron scattering, however, this has
to be taken into account. The observable charge density
is calculated by convolution of the proton density with a Gaussian form
factor~\cite{Negele70} with a proton size $a = \sqrt{2/3} \;
\langle r^2 \rangle^{1/2}_p = 0.65$~fm, which for spherically
symmetric density distributions leads to
\begin{equation}
\label{charge-sph}
\rho_{\rm ch} (r)
  =
    \frac{1}{a\sqrt{\pi}}
    \int \! dr' r' \, \rho_p(r') \,
    \left[
      \frac{e^{-(r-r')^2/a^2}}{r} - \frac{e^{-(r+r')^2/a^2}}{r}
    \right] \, .
\end{equation}
The charge density (right panel)  is compared in Figure~\ref{charge-dens} to
the point proton density (left panel) for the same four cases discussed in
Fig.~\ref{scaled-dens}. Like correlations, the convolution~(\ref{charge-sph})
tends to even out the variations of the density profile: the central density
rises and the maximum density of the outer bulge becomes smaller.
In combination, both leads to a substantial reduction of the depletion
factor from $F_{\text{max}}=0.41$ for the point proton density in a
spherical HF calculation to $F_{\text{max}} = 0.09$ for the charge density
of the $0^+$ GCM ground state.

Adding correlations, the root-mean-square (rms) radius of the point proton
density increases from 3.127~fm for the spherical HF state to 3.133~fm for
the particle-number projected spherical HFB+LN state and to 3.180~fm for
the GCM $0^+$ ground state. Looking at the density profiles in
Fig.~\ref{charge-dens} the larger radius of the GCM $0^+$
ground state might appear counter-intuitive, as, at small radii $< 3$~fm,
the protons are obviously shifted to the inside. At larger radii,
however, the tail of the density of the GCM $0^+$ ground state
becomes slightly larger than the density of the other states,
which is almost undetectable on the linear scale of Fig.~\ref{charge-dens}.
Because of the factor $r^4$ in the mean-square radius integral in
polar coordinates, this tail is much more important than the
center of the nucleus.

Figure.~\ref{charge-dens} puts into evidence that the reduction of
the depletion factor at each stage of the calculation is partly due
to the reduction of the maximum density $\rho_{\text{max}}$
Obviously, shell effects can reduce the density at some radii, but
also enhance it at others. This indicates that the definition of the
depletion factor~(\ref{eq:F}) contains an ambiguity concerning the
reference density. An alternative definition of a depletion factor
could be
\begin{equation}
\label{eq:F:sat}
F_{\text{sat},\tau}
\equiv \frac{ \rho_{\textrm{sat},\tau} - \rho_{\textrm{cent},\tau}}
            {\rho_{\textrm{sat},\tau}}
\, ,
\end{equation}
where $\rho_{\textrm{sat},\tau}$ with $\tau=p$, $n$, $t$ is now the
saturation value of the proton, neutron, and total density.
For $\elemA{34}{Si}$, we have $\rho_{\text{sat},p} = (14/34) \times
0.16$~fm$^{-3} = 0.066$~fm$^{-3}$, $\rho_{\text{sat},n} = (20/34) \times
0.16$~fm$^{-3} = 0.094$~fm$^{-3}$, and $\rho_{\text{sat},t} = 0.16$~fm$^{-3}$,
respectively. Unlike $F_{\text{max}}$, this alternative depletion factor
$F_{\text{sat}}$ can also be used to quantify central bumps in the
density distribution.

\begin{table}[t!]
\caption{
\label{tab:F:proton}
Central and maximum proton density of $\elemA{34}{Si}$ and the
depletion factors $F_{\text{max},p}$~[cf.\ Eq.~(\ref{eq:F})] and
$F_{\text{sat},\tau}$~[cf.\ Eq.~(\ref{eq:F:sat})].
For the latter, values for proton, neutron and the total
densities are given. All densities are in fm$^{-3}$. The three
first lines correspond to a spherical state. The values
labeled with $N\&Z$, $J=0$ correspond to the prolate
minimum of the $N\&Z$, $J=0$ projected energy curve
of Fig.~\ref{energy_curves}.
}
\begin{center}
\begin{tabular}{lcccccc}
\hline \noalign{\smallskip}
     &
$\rho_{\text{cent},p}$ &
$\rho_{\text{max},p}$  &
$F_{\text{max},p}$     &
$F_{\text{sat},p}$   &
$F_{\text{sat},n}$   &
$F_{\text{sat},t}$       \\
\noalign{\smallskip} \hline \noalign{\smallskip}
HF              & 0.044 & 0.074  & 0.41  &  0.34 & $-0.37$ & $-0.08$ \\
HFB+LN          & 0.050 & 0.074  & 0.32  &  0.24 & $-0.31$ & $-0.08$ \\
$N$\&$Z$        & 0.047 & 0.074  & 0.36  &  0.28 & $-0.30$ & $-0.06$ \\
$N$\&$Z$, $J=0$ & 0.051 & 0.073  & 0.30  &  0.22 & $-0.27$ & $-0.07$ \\
GCM(g.s.)       & 0.057 & 0.073  & 0.21  &  0.13 & $-0.22$ & $-0.07$ \\
\noalign{\smallskip} \hline
\end{tabular}
\end{center}
\end{table}

\begin{table}[t!]
\caption{
\label{tab:F:charge}
Same as Table~\ref{tab:F:proton}, but for the charge density,
Eq.~(\ref{charge-sph}).}
\begin{center}
\begin{tabular}{lcccc}
\hline \noalign{\smallskip}
     &
$\rho_{\text{cent}}$ &
$\rho_{\text{max}}$  &
$F_{\text{max}}$     &
$F_{\text{sat}}$       \\
\noalign{\smallskip} \hline \noalign{\smallskip}
HF               & 0.056 & 0.071  & 0.21 & 0.15 \\
HFB+LN           & 0.060 & 0.071  & 0.16 & 0.09 \\
$N$\&$Z$         & 0.058 & 0.071  & 0.18 & 0.12 \\
$N$\&$Z$, $J=0$  & 0.060 & 0.070  & 0.14 & 0.09 \\
GCM(g.s.)        & 0.064 & 0.070  & 0.09 & 0.04 \\
\noalign{\smallskip} \hline
\end{tabular}
\end{center}
\end{table}

The results of Figs.~\ref{scaled-dens} and~\ref{charge-dens} using
these two depletion factors are summarized in Tables~\ref{tab:F:proton}
and \ref{tab:F:charge}. The value of $\rho_{\textrm{sat}, p}$ is
systematically smaller than the values of $\rho_{\textrm{max},p}$ and,
as a consequence, the values of $F_{\textrm{sat}}$ for protons are
smaller than those of $F_{\text{max}}$. For neutrons, the value of
$F_{\text{max},n}$ is always negative because of the central bump
of the neutron density distribution. Again, the large central
bump predicted by the HF calculation is reduced when
correlations are added. Altogether, the central total density is
always larger than saturation density as evidenced by $F_{\text{sat},t}$.


\section{Further discussion and conclusions}
\label{concl}

There are two major differences between the structure of $\elemA{34}{Si}$
as predicted by our calculation and the structure of other candidates for
bubble structure that have been discussed in the 1970s-1990s. First, as
discussed above, the depletion of the central density in $\elemA{34}{Si}$
appears for the proton density only, whereas the total density has an
almost flat distribution throughout the bulk of the nucleus.
Second, the level ordering of bubble nuclei is usually different from the
one of regular nuclei, which is not the case for $\elemA{34}{Si}$. Take,
for example, the hypothetical bubble-type configuration of the
$\elemA{36}{Ar}$ discussed in Ref.~\cite{Davies73}. There, the sequence
of single-particle levels is altered such that the $2s_{1/2}$ level is
pushed above the $1d_{3/2}$ level for both protons and neutrons. By
contrast, for $\elemA{34}{Si}$ only the relative distance of levels
is changed such that the $Z=14$ gap opens up, cf.~Fig.~\ref{spe}. In
the absence of a bubble structure of the total density, and of the
rearrangement of shells that is typical for bubble nuclei, the predicted
anomaly of the proton density distribution of $\elemA{34}{Si}$ appears
to be an example of a central depression of the density, as
observed also for many other nuclides~\cite{Friedrich86}, rather than a nuclear bubble.

Taking into account correlations reduces the central depletion of the
proton density in $\elemA{34}{Si}$, as expected from earlier studies of
other systems. Our main findings are:
\begin{enumerate}
\item[(i)]
A HFB+LN calculation overestimates proton pairing correlations in
$\elemA{34}{Si}$. Particle-number projection of the spherical state
constructed with HFB+LN reduces the pairing correlations, such that
the density profile almost goes back to the HF one. Clearly, a treatment
of pairing correlations beyond the mean field with exact particle-number
projection is needed in this case.
\item[(ii)]
Fluctuations in quadrupole degrees of freedom strongly even out the
fluctuations in the density profile; the central depression and the
outer bump of the proton density are reduced, as is the central bump
of the neutron density.
\item[(iii)]
The central depletion of the density is less pronounced
when looking at the experimentally observable charge density instead
of the point proton density.
\end{enumerate}
While all of the above findings can be expected to be generic on a
qualitative level, the quantitative increase of the depletion factor
when going from  spherical HF to full projected GCM might depend on
choices made for the effective interaction. In particular, in view of
the somewhat too large mixing that we find between the two lowest $0^+$
GCM states, our calculation  might slightly overestimate the role of
shape fluctuations in the ground state.

As discussed in the introduction, a central depletion of the proton density of $\elemA{34}{Si}$ has been suggested as an explanation for the reduction by about 0.6~MeV of the spin-orbit splitting of the neutron $3/2^-$ and $1/2^-$ levels inferred from transfer reactions~\cite{Burgunder2011,Sorlin2011}. One has to be careful about such conclusion. First, the connection between the centroids of the spectral strength function of combined one-nucleon pick-up and removal reactions and the eigenvalues of the single-particle
Hamiltonian is model-dependent~\cite{Duguet11}, and when looking at the dominating fragments only, the comparison is far from clear. Second, as the symmetry-restored GCM method corresponds to a superposition of many states obtained with different mean fields, there is no straightforward
procedure that would allow for a statement about effective single-particle energies based on the density profile from our calculation.
The only meaningful way to compare with the data would be to perform the same kind of calculation for \elemA{35}{Si} and \elemA{37}{S}, which at present, however, is out of our reach.

When discussing the spin-orbit splitting of the neutron $2p$ levels in $\elemA{34}{Si}$, there is an additional complication that goes even
beyond these considerations. At spherical shape, the neutron $2p_{3/2}$
and $2p_{1/2}$ levels are far above the Fermi energy and outside of the
energy interval shown in the Nilsson diagram of Fig.~\ref{spe}. In our
spherical HF calculation of $\elemA{34}{Si}$, the former is weakly bound
at $-0.56$~MeV, whereas the latter is even unbound at $+0.83$~MeV.
In such a situation, the coupling to the continuum has to be carefully
taken into account, which is a task that goes beyond our study.
By contrast, both single-particle levels are predicted
to be (weakly) bound in a similar HF calculation for $\elemA{36}{S}$.

In summary, we find that correlations from pairing and fluctuations in
quadrupole deformation substantially reduce the central depletion of
the proton density in $\elemA{34}{Si}$. The extension of our method
to the calculation of transition densities in the laboratory frame
as observables in inelastic electron scattering is currently
underway~\cite{Yao2012}.

%
\section{Acknowledgments}

Fruitful discussions with Stephane Gr{\'e}vy, Witold Nazarewicz and
Olivier Sorlin are gratefully acknowledged.
This research was supported in parts by
the PAI-P6-23 of the Belgian Office for Scientific Policy,
the F.R.S.-FNRS (Belgium),by the European Union's Seventh Framework Programme under grant agreement n°262010,
the National Science Foundation of China under Grants No.~11105111 and
No.~10947013,
the Fundamental Research Funds for the Central Universities (XDJK2010B007),
the Southwest University Initial Research Foundation Grant to Doctor
(SWU109011),
the French Agence Nationale de la Recherche under Grant
No.~ANR 2010 BLANC 0407 "NESQ",
and by the CNRS/IN2P3 through the PICS No.~5994.

%
%

\end{document}